\documentclass[showpacs,floatfix,twocolumn,prb,superscriptaddress]{revtex4-1}

\usepackage{graphicx,color}
\usepackage{amsmath}

\begin{document}

\title{Uncovering anisotropic magnetic phases via fast dimensionality analysis}

\author{Manohar H. Karigerasi}
\affiliation{Materials Science and Engineering Department, University of Illinois Urbana-Champaign, IL 61801, USA}

\author{Lucas K. Wagner}
\affiliation{Physics Department, University of Illinois Urbana-Champaign, IL 61801, USA}

\author{Daniel P. Shoemaker}\email{dpshoema@illinois.edu}
\affiliation{Materials Science and Engineering Department, University of Illinois Urbana-Champaign, IL 61801, USA}


\begin{abstract}
A quantitative geometric predictor for the dimensionality of magnetic interactions is presented. 
This predictor is based on networks of superexchange interactions and can be quickly calculated for crystalline compounds of arbitrary chemistry, occupancy, or symmetry. 
The resulting data are useful for classifying structural families of magnetic compounds. 
We have examined compounds from a demonstration set of 42,520 materials with $3d$ transition metal cations. 
The predictor reveals trends in magnetic interactions that are often not apparent from the space group of the compounds, such as triclinic or monoclinic compounds that are strongly 2D. 
We present specific cases where the predictor identifies compounds that should exhibit competition between 1D and 2D interactions, and how the predictor can be used to identify sparsely-populated regions of chemical space with as-yet-unexplored topologies of specific $3d$ magnetic cations.
The predictor can be accessed for the full list of compounds using a searchable frontend, and further information on the connectivity, symmetry, valence, and cation-anion and cation-cation coordination can be freely exported.

\end{abstract}

\maketitle

\section{Introduction} 

There are a number of unique properties that require or are correlated with specific topologies of magnetic interactions. 
For example, hexagonal, kagom\'{e}, or pyrochlore lattices of spins are predicted to form classical spin glasses or a quantum spin liquids.\cite{balents_spin_2010} 
Layered oxides, chalcogenides, and pnictides based on iron and copper have two-dimensional square planar lattices of spins that can be doped to produce complex phase diagrams including high-temperature superconductivity.\cite{Sokol1993,Magalhaes2012,Luetkens2009,Paglione2010}
It is a major goal of materials science to find new materials with these motifs, compare interactions in compounds comprised of similar motifs, and to find new magnetic motifs that have yet-unknown properties.

Magnetic motifs are not necessarily easily found by using standard crystallographic descriptors such as space groups and structure types. 
This is in contrast to polar materials (centers of inversion) or antiferromagnets with accessible current-driven spin torques (global and site symmetry).\cite{halasyamani_noncentrosymmetric_2010,zhang_hidden_2014}
For example, the superconducting parents La$_2$CuO$_4$ and BaFe$_2$As$_2$ are accompanied by many lower-symmetry compounds with similar motifs of square-planar Cu--O and square nets of Fe--$Q$ tetrahedra where $Q$ is a pnictogen or chalcogen. Bi$_2$Sr$_2$CaCu$_2$O$_{8+\delta}$, and (CaFe$_{1-x}$Pt$_x$As)$_{10}$Pt$_{4-y}$As$_8$ contain familiar CuO$_4$ and FeAs$_4$ motifs, but the former contains modulations that induce a monoclinic distortion, and the latter is triclinic.\cite{Lohnert2011,kakiya_superconductivity_2011,gladyshevskii_modulated_1996}
These compounds are pseudo-two-dimensional, in that spin correlations persist far above room temperature in the Cu--O or Fe--As layers, but correlation between layers is relatively weak.\cite{pickett_electronic_1989,matan_anisotropic_2009,mcqueeney_anisotropic_2008}

Performing a motif-based search for compounds manually is common, but slow and unreliable.\cite{dshemuchadse_structural_2011,cava_crystal_2013,hosono_iron-based_2015,bradlyn_beyond_2016}
We seek a fast, robust algorithm to quantify the dimensionality of magnetic interactions with specific connectivities.
Several previous studies have used high-throughput methods to classify 2-dimensional materials (irrespective of magnetism) that are separated by van der Waals bonding between the layers.
Klintenberg, \textit{et. al.} imposed geometric constraints (trigonal, hexagonal or tetragonal crystal systems, a low atomic packing fraction, and gaps along one of the crystal axes) to produce a list of 2D materials and their electronic band structures. \cite{klintenberg_data_mining_2013}
Such a search does not accomplish our goal of analyzing pseudo-2D magnetic materials, since our compounds of interest often have low symmetry ($P\overline{1}$ CdCu$_3$(OH)$_6$(NO$_3$)$_2\cdot$H$_2$O) and strong interlayer bonding (BaFe$_2$As$_2$, La$_2$CuO$_4$). 
Other methods based on iterative slab generation are too computationally demanding to screen through many complex structures.\cite{prashun_slab_structure_2016}

Here we build on a relation described by Mounet, \textit{et al.}\cite{marzari_method_2016} to categorize van der Waals-layered materials.
We create an iterative framework to generate a quantitative geometric predictor of quasi-0-, 1-, 2-, or 3-dimensionality in magnetic materials. 
The cation-anion and cation-cation connectivities of all layers and chains are interrogated quickly and the full information is returned in a searchable database.
Information about the connectivity of the metal ions (superexchange pathways)  can serve as a powerful classification scheme for materials with similar magnetic motifs but disparate space group symmetry.

\begin{figure}
\centering\includegraphics[width=0.85\columnwidth]{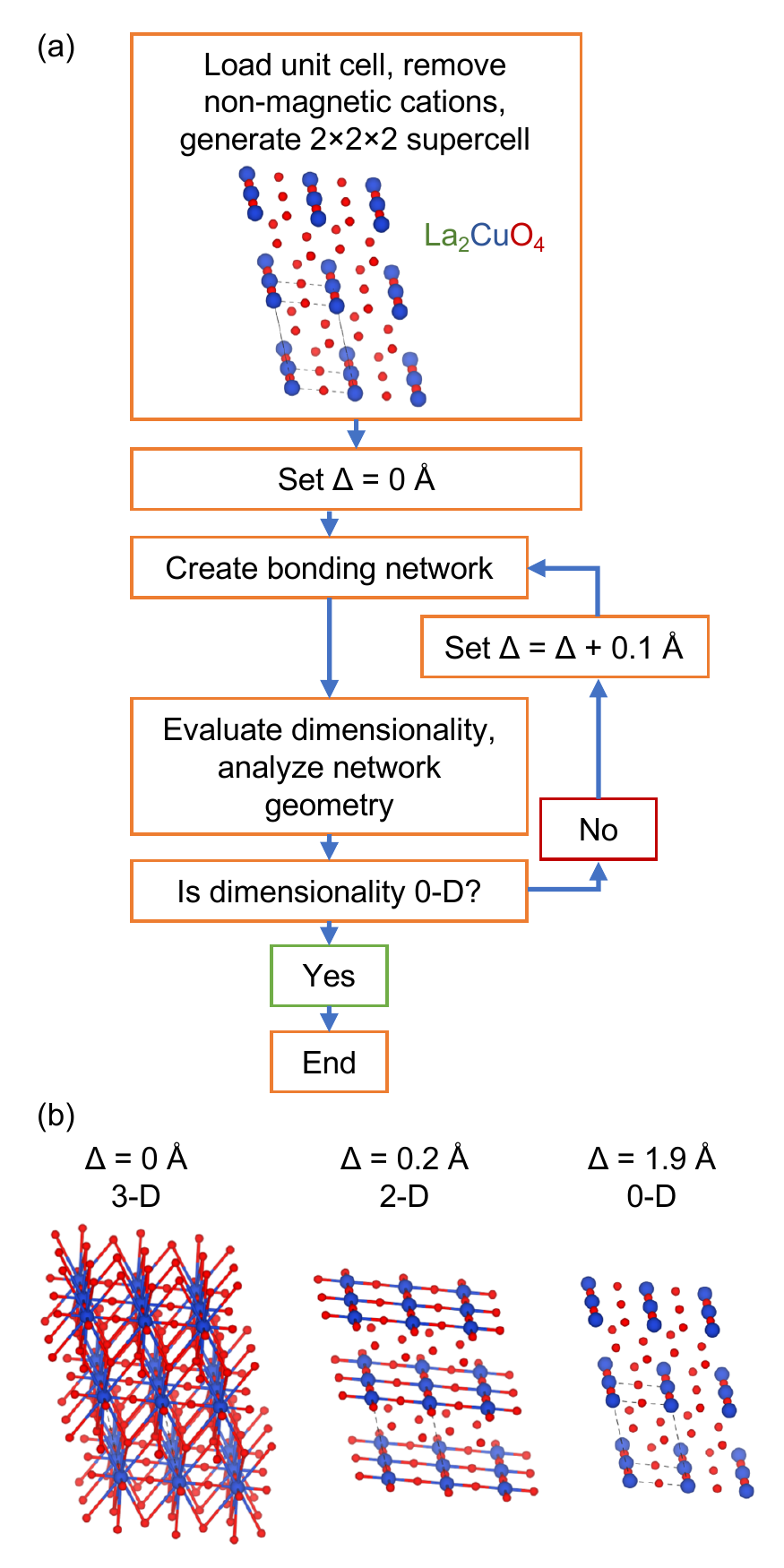} \\
\caption{In (a), the method for quantifying magnetic dimensionality is shown to iterate through increasing values of $\Delta$ until the bonding network between magnetic ions is disconnected. This progression is shown schematically in (b) for the case of the Cu-O sublattice in La$_2$CuO$_4$.}
\label{fig:flowchart}
\end{figure}

\section{Methods}

Our test set is composed of materials that can exhibit 3$d$ magnetism. 
Compounds from the Inorganic Crystal Structure Database (ICSD version 3.5.0)\cite{Hellenbrandt_icsd_2004}  are considered if they contain (a) at least one commonly magnetic cation in the form of V, Cr, Mn, Fe, Co, Ni or Cu, (b) at least one anion defined as C, pnictogens, chalcogens or halogens, 
and (c) any number of other atoms.
Structures with cations with partially-filled $4d$, $5d$, $4f$, or $5f$ orbitals are included here, but the magnetic dimensionality of the heavier cation is not examined; it is straightforward to add them to the list of allowed magnetic cations in the algorithm. 
Likewise, an equivalent classification may be performed for intermetallics without modification of the algorithm.
Compounds exhibiting exactly the same chemical formula and space groups as other compounds are tagged as duplicates. 
Excluding duplicates, our set is 42,520 compounds. 
We have not placed any restrictions on partial occupancies, hydrated materials, or symmetry. 
Structural manipulations are performed using \texttt{pymatgen},\cite{Shyue_pymatgen_2013} and bonding networks are stored and evaluated using \texttt{networkx}.\cite{hagberg_networkx_2008}

\paragraph{Constructing the bonding network.} The algorithm is performed as shown in the flow chart of Figure \ref{fig:flowchart}. 
First, non-magnetic cations are deleted from the structure, followed by the creation of a $2 \times 2 \times 2$ superlattice. 
Deleting non-magnetic cations is not strictly required, but accelerates the algorithm since these cations are not considered part of the magnetic superexchange networks. 
Bonds are formed between pairs of ions that are closer than a cutoff distance R$_{cut}$ :
\begin{equation}
R_{cut} = R_1 + R_2 - \Delta
\end{equation}
where R$_1$ and R$_2$ are the Van der Waals radii\cite{Alvarez2013} of ion 1 and ion 2 respectively and $\Delta$ is some constant greater than 0~\AA\ and less than 3~\AA, a distance larger than any bonded ion radius in the system.
We take the occupancy-weighted mean of radii $R_i$ for ions that share crystallographic sites (Vegard's law).\cite{vegard_konstitution_1921,vegard_xv._1928}
After drawing bonds for a given $\Delta$, vectors are connected from each unique magnetic site to its periodic images in neighboring unit cells. Vectors that point to disconnected bonding networks are discarded, and the dimensionality is found by taking the rank of the remaining vectors. 

\paragraph{Quantifying dimensionality.} 
Compounds that exhibit 1- or 2-dimensionality for any value of $\Delta$ are assigned a predictor $\Delta_\text{range}$, which is the difference between the highest ($\Delta_\text{max}$) and lowest ($\Delta_\text{min}$) values that return a magnetic network with a given dimensionality. Compounds that exhibit both 1D and 2D character have two values of $\Delta_\text{range}$: one for 1D and one for 2D. All compounds must become 3D if $\Delta$ is small (or negative) and must become 0D if $\Delta$ is very large.

\paragraph{Database organization.} 
The results can be searched, filtered, and output using a web-based frontend accessible from the Illinois Data Bank.\cite{karigerasi_geometric_2018}
Stored properties of each compound include their 1-D and 2-D characters, cations and anions in the magnetic network, chain and plane directions, metal-anion and metal-metal coordination and angles, and ICSD-assigned values (formula, space group, cell volume, etc.).
All compounds were tagged with their ICSD-assigned structure type. 
Compounds without an assignment were tagged if they matched the exact formula and space group of a previously-tagged compound, and the remaining compounds were assigned a new structure type if they did not match any assigned compounds using the \texttt{StructureMatcher} class of \texttt{pymatgen} with default tolerances.


\section{Results and Discussions}

\subsection{Understanding $\Delta$ and examining common motifs}

Most uses of the dimensionality predictor are guided or filtered by constraints on three parameters: $\Delta_\text{range}$, cation-anion coordination, and cation-cation coordination, which are all evaluated in the formation of bonding networks. We describe the parlance and behavior of these parameters here, then give examples of useful dimensionality searches.

\begin{figure}
\centering\includegraphics[width=0.85\columnwidth]{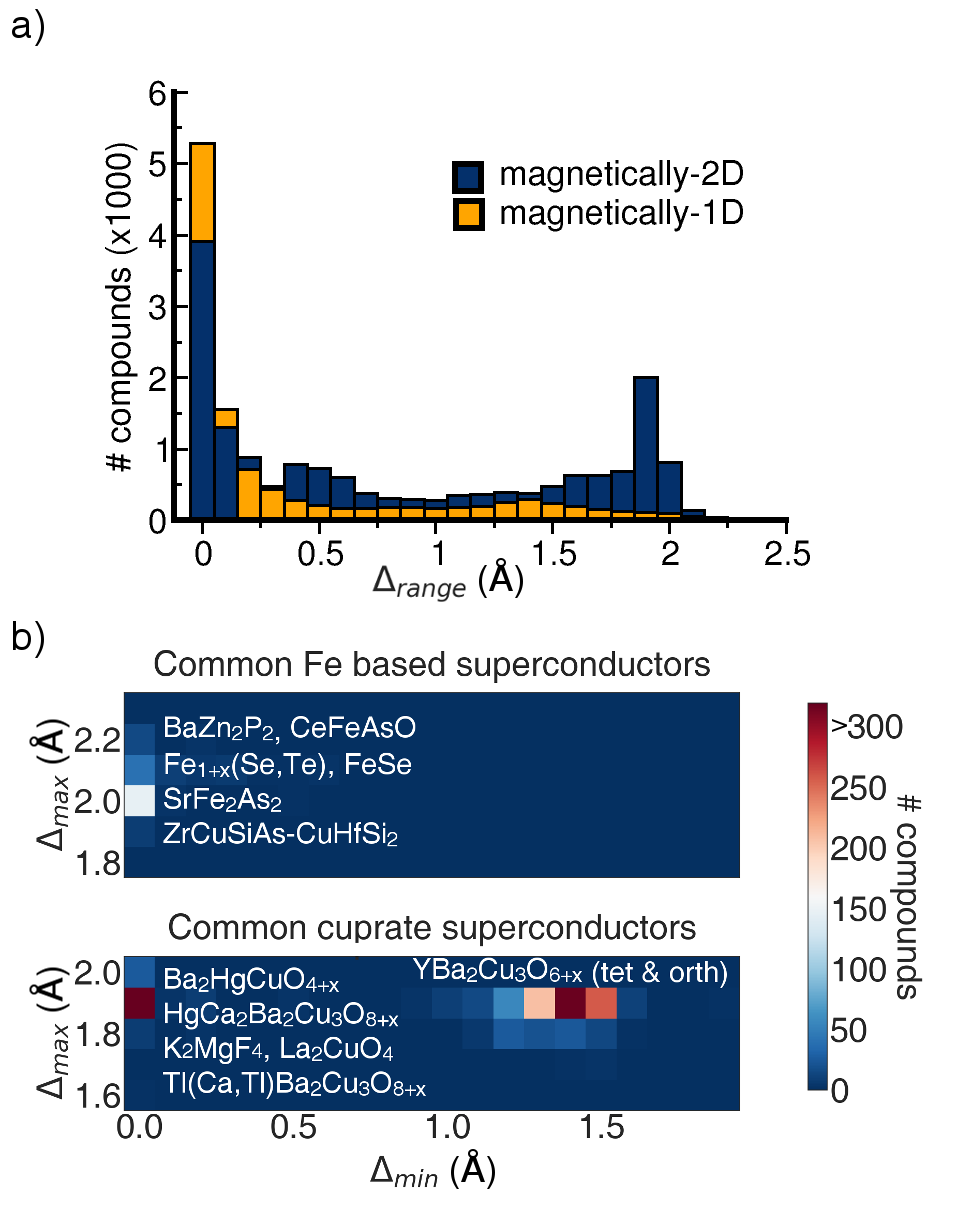} \\
\caption{The $\Delta_\text{range}$ distribution of magnetically-1D and magnetically-2D compounds is shown in (a).  
The maximum and minimum values of $\Delta$ that return 2D connectivity for common Fe-based and cuprate superconductors, spanning many structure types, are shown in (b). Most are tightly clustered except for YBCO-type compounds, which have  $\Delta_\text{min} = 1.5$ and $\Delta_\text{max} = 1.9$~\AA\ for a relatively small 2D $\Delta_\text{range} = 0.4$~\AA. 
} 
\label{fig:delta_range}
\end{figure}

The \textbf{geometric dimensionality predictor $\Delta_\text{range}$} is best understood in the context of known compounds.
Compounds that would be easily identified (\emph{i.e.}, visually) as magnetically two-dimensional have $\Delta_\text{range}$ larger than the bonding radii of the material.
For example, the van der Waals compound CrI$_3$ and layered delafossite NaCrO$_2$ have $\Delta_\text{range}$ = 1.7 \AA\ and 1.8 \AA, respectively, so the layers must be well-spaced.
Figure \ref{fig:delta_range}(a) shows the distribution of $\Delta_\text{range}$ for compounds that return a nonzero $\Delta_\text{range}$ for 1- and 2-dimensionality. 
From this histogram, it is clear why magnetic dimensionality requires a quantitative predictor, rather than a boolean: both distributions are smooth and bimodal, with peaks near $\Delta_\text{range}$ = 0~\AA,  1.5~\AA\ for 1D, and 1.9~\AA\ for 2D. 
Compounds with well-spaced 2D sheets and 1D chains fall into the latter two distributions, while compounds that are better described as a different dimensionality lie toward $\Delta_\text{range} = 0$~\AA.
Such compounds that are poorly described as 1D or 2D generally have weakly anisotropic structures, such as V$_3$Se$_4$, which is effectively 3D and shows long range antiferromagnetic order below 16~K.\cite{Kallel1984,Kitaoka1979}
Likewise, monoclinic ScMnO$_3$ is a perovskite with 3D connectivity which is ferromagnetic below 100~K and antiferromagnetic below 51~K.\cite{doi:10.1021/ic4016838,2014JPCM...26W5402Y} 
Both have a small 2D $\Delta_\text{range} = 0.1$~\AA.

\begin{figure}
\centering\includegraphics[width=0.85\columnwidth]{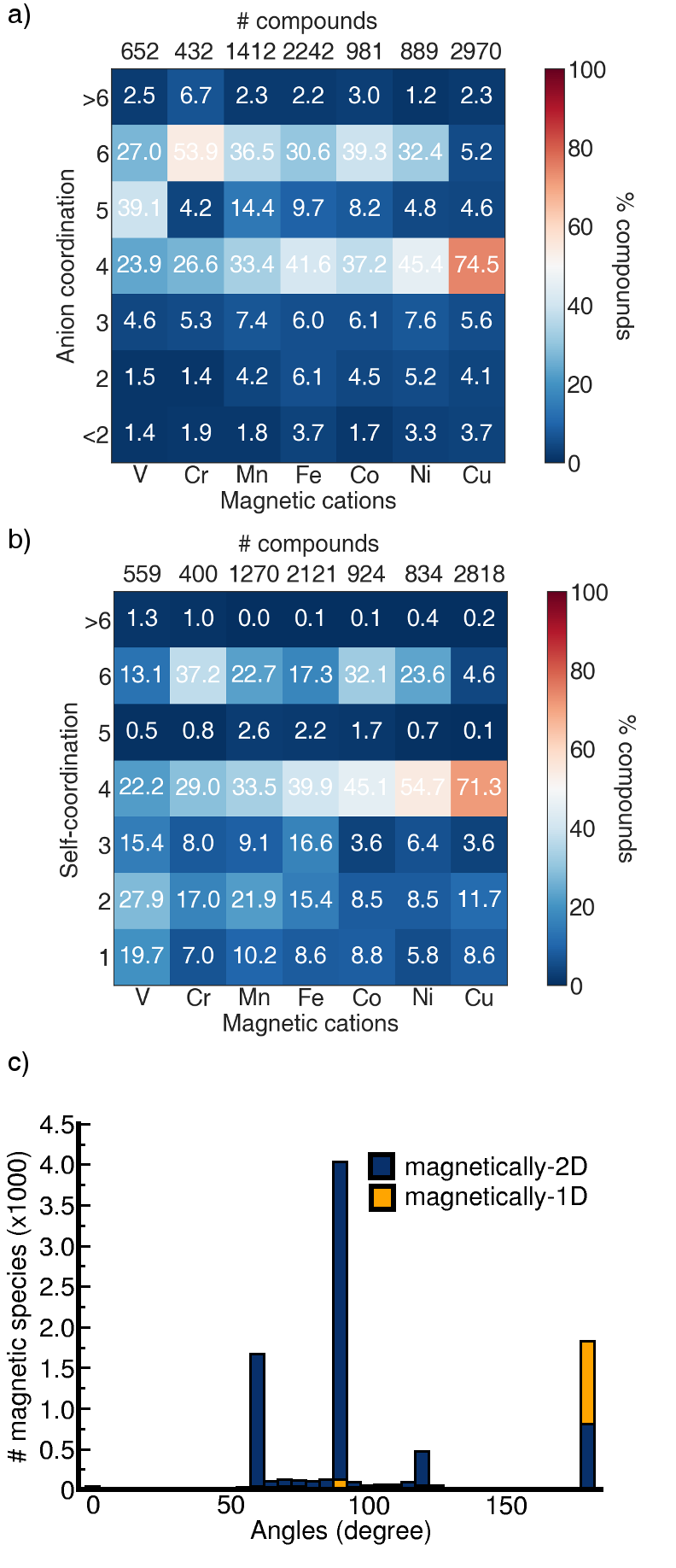} \\
\caption{Coordination networks for magnetic cations with (a) anions and (b) cations themselves are extracted during the evaluation of dimensionality. The angles formed between magnetic cations in the network are shown in (c). Compounds having both magnetic 2D and 1D character were assigned a dimension based on their higher $\Delta_\text{range}$ to avoid double counting.} 
\label{fig:coordination}
\end{figure}

The distribution of \textbf{anion coordination of magnetic cations} is shown in Figure \ref{fig:coordination}(a), taken at $\Delta_\text{max}$ for each compound with 2D $\Delta_\text{range} \geq 0.8$~\AA.
Cr, Mn, and Co are most commonly 6-fold coordinated (octahedral mostly), while Fe, Ni, and Cu are most commonly 4-fold (square planar or tetrahedral).
The latter case is biased by the over-reporting of unconventional superconductors. 
A peak in 5-coordination for V highlights its tendency to form square pyramidal units with O, for example in Sr$_2$VO$_3$FeAs (an unconventional superconductor),\cite{Zhu2009} CaV$_2$O$_5$ (a spin-1/2 ladder vanadate),\cite{Korotin2000} Rb$_2$V$_3$O$_8$ (a square lattice spin-1/2 Heisenberg antiferromagnet),\cite{2006PhRvB..74u4424L} etc.

Examining \textbf{the cation-cation connectivity} (a predictor of superexchange pathways) requires next-nearest-neighbor distances, which are formed in our dimensionality analysis but not available as a search criterion in other structural databases. 
This self-coordination of magnetic cations is evaluated for nearest self-neighbors (with tolerance 0.1~\AA) in a single bonding network.
For magnetically-2D compounds, the relative prevalence of metal-metal coordination number for each magnetic cation is shown in Figure \ref{fig:coordination}(b). 
The intralayer metal-metal coordination number of 4 (square, diamond, and rectangle) is most common for Fe, Co, Ni and Cu, while Cr is most commonly found in 6-fold hexagonal patterns. 
Compounds in Figure \ref{fig:coordination}(b) where the cation-cation coordination is lower than 4 (\emph{e.g.} 2-fold V) typically exhibit multiple different distances in the layer, typical of monoclinic, orthorhombic or triclinic layered structures.
A histogram of the cation-cation-cation angle distribution has been plotted in \ref{fig:coordination}(c).
The 60$^\circ$ magnetically-2D compounds contain hexagonal or kagom\'{e} layers and 180$^\circ$ magnetically-2D compounds consist primarily of orthorhombic, monoclinic or triclinic crystal systems where the nearest self-neighbors lie in a straight line. 
A small number of 1D compounds at 90$^\circ$ represent spin-ladder and ribbon compounds (BaFe$_2$Se$_3$,\cite{2012PhRvB..85r0405C}, K$_3$Cu$_3$Nb$_2$S$_8$,\cite{LU1992312} etc.).

\begin{figure}
\centering\includegraphics[width=0.85\columnwidth]{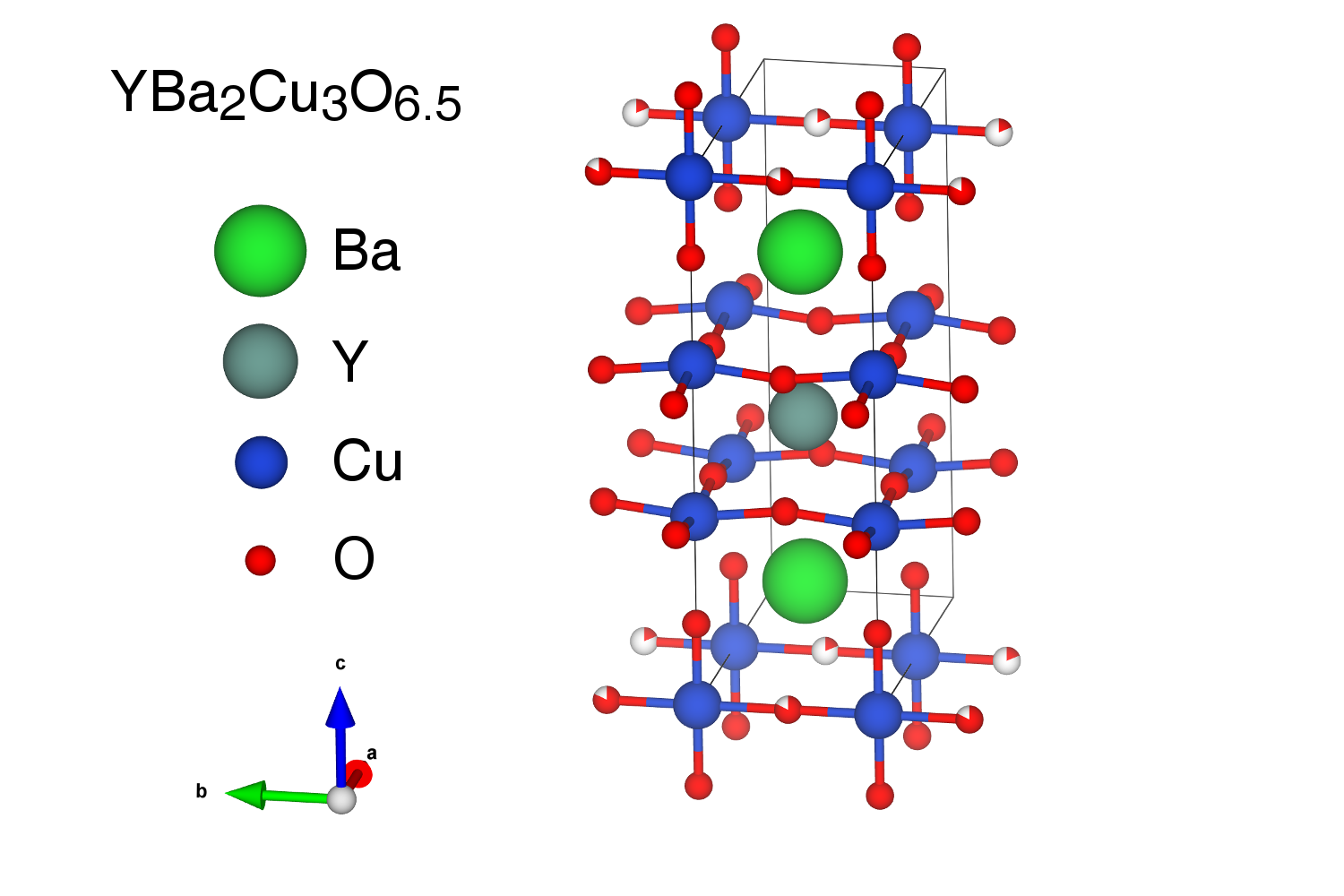}\\
\caption{
Unlike most other cuprate superconductors, YBa$_2$Cu$_3$O$_{6.5}$ has a small 2D $\Delta_\text{range} = 0.4$~\AA\ since the layer separation and hence Cu--O distance in the $c$ direction are comparatively small.} 
\label{fig:YBCO}
\end{figure}

With dimensionality, anion coordination, and cation connectivity in hand, it is straightforward to show concrete examples to quickly extract specific families of functional materials:
\begin{enumerate}

\item
\textbf{Unconventional iron-based superconductors} populate searches for Fe with Se or As with Fe-Fe-Fe angles between 85-95$^\circ$ and $\Delta_\text{range} \geq 2.0$~\AA.
This search returns 274 compounds, which includes BaFe$_2$As$_2$-related phases, including FeSe, LaFeOAs, LiFeAs, K$_x$Fe$_{2-y}$Se$_2$, (CaFe$_{1-x}$Pt$_x$As)$_{10}$Pt$_{4-y}$As$_8$, etc. The corresponding 2D $\Delta_\text{min}$ and $\Delta_\text{max}$ values for those of common structure types are shown in  Figure \ref{fig:delta_range}(b). Despite different distortions and space groups, they exhibit a very narrow spread of $\Delta_\text{range}$.

\item
\textbf{Cuprate superconductors} are readily accessed by a search for Cu and O with Cu-Cu-Cu angles of 85-95$^\circ$ and $\Delta_\text{range} \geq 1.9$~\AA, which returns a list of 1136 compounds of families from La$_2$CuO$_4$, the Bi-, Tl- and Hg- based cuprates, regardless of their layering symmetry or space group. The distribution of 2D $\Delta$ values for most common cuprates is tightly clustered and shown in Figure \ref{fig:delta_range}(b).
Cuprates in the YBa$_2$Cu$_3$O$_{7-x}$ (YBCO) family, on the other hand, are shown in Figure \ref{fig:delta_range}(b) to have a lower 2D $\Delta_\text{range} = 0.4$~\AA\ due to close proximity of Cu$^{2+}$ along the stacking direction shown in Figure \ref{fig:YBCO}.
The relatively close interlayer spacing in YBCO compounds is reflected in the N\'{e}el temperature of the parent, which can exceed 500~K.\cite{tranquada_antiferromagnetism_1988}
In fact, this strong interlayer coupling hints at the fact that unadulterated spin-1/2 Heisenberg layers are poorly manifested in the cuprates,\cite{manousakis_spin-_1991,plakida_high-temperature_2010} but they do have an associative relationship so the motifs aid in identifying such compounds.

\item
\textbf{Quantum spin liquid} candidates are best found by searching for compounds having $\Delta_\text{range} \geq 1.0$~\AA\ and Cu-Cu-Cu angles from 55 to 65$^\circ$. We believe this search returns a \emph{complete list} of quantum spin liquids that have been discovered to date. Specifying the metal-metal coordination enables discrimination between hexagonal (6-coordinate, e.g. $R$CuO$_2$ delafossites where $R$ = Rh, Al, In, Sc, Y, Ga and Pr)\cite{olariu_sr_2006,garlea_incuo_2004} and kagom\'{e} compounds (4-coordinated, e.g. BaCu$_3$V$_2$O$_8$(OH)$_2$-like  $\beta$-vesignieites,\cite{Yoshida2012} herbertsmithites and kapellasites such as ZnCu$_3$(OH)$_6$Cl$_2$, etc.).\cite{Yoshida2012,Shores2005}

\item
\textbf{Quasi-1D spin-1/2 Heisenberg cuprate antiferromagnets} can be filtered by selecting 1D compounds with Cu--O bonding and the desired Cu--Cu coordination, typically 2-fold for single chains. Limiting to square-planar CuO$_4$ chains returns compounds such as Li$_2$CuO$_2$, which exhibits temperature-independent susceptibilities at low temperatures.\cite{Sapina1990} 
Compounds with corner-sharing cuprate units also appear, such as Sr$_2$CuO$_3$, which has high-temperature magnetic susceptibility that approximates a spin-1/2 Heisenberg antiferromagnetic extremely well, with $J > 2000$~K.\cite{Motoyama1996} 
Sr$_2$CuO$_3$ has an appreciable 2D $\Delta_\text{range} = 1.6$~\AA\ since the inter-chain distance between the Cu$^{2+}$ ions is relatively small, about 3.5~\AA.
Some examples of compounds that contain isolated CuO$_4$ ions separated by a different polyhedra uncovered by this search include K$_2$CuP$_2$O$_7$ and Sr$_2$Cu(PO$_4$)$_2$. 
Both of these compounds contain CuO$_4$ units connected by PO$_4$ tetrahedra and are excellent manifestations of spin-1/2 Heisenberg antiferromagnetic chains.\cite{Nath2008,belik_characterization_2004} 
\end{enumerate}

\subsection{Uncovering unconventional compounds}

\begin{figure}
\centering\includegraphics[width=\columnwidth]{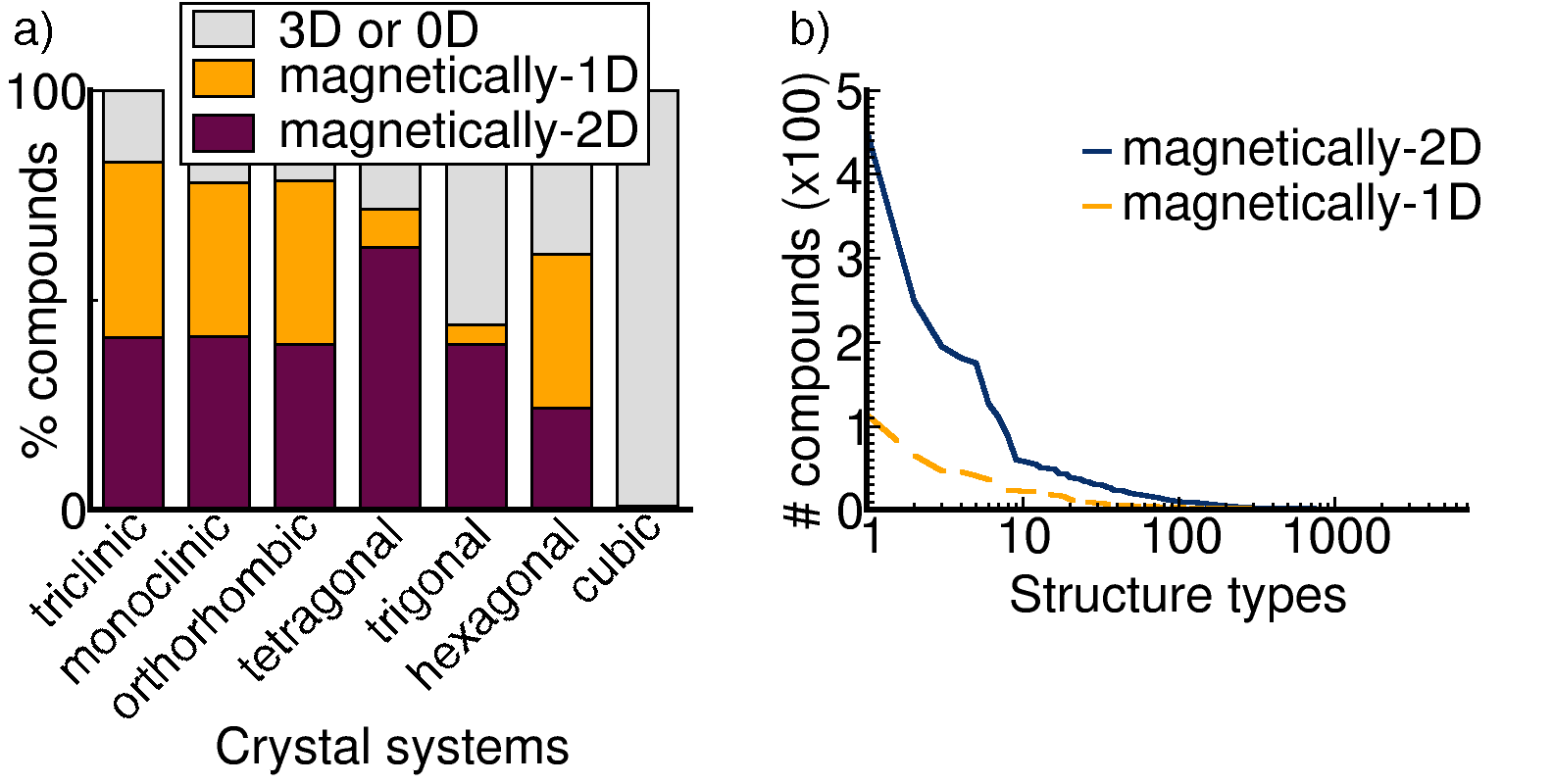} \\
\caption{ The distribution of magnetically 1D and 2D compounds into crystal systems is shown in (a). 
In (b), the distribution of compounds into structure types reveals that a relatively small number of structure types are well-populated, out of a total 6403 2D and 5268 1D structure types.}
\label{fig:system_stype}
\end{figure}

\begin{figure*}
\centering\includegraphics[width=0.85\paperwidth]{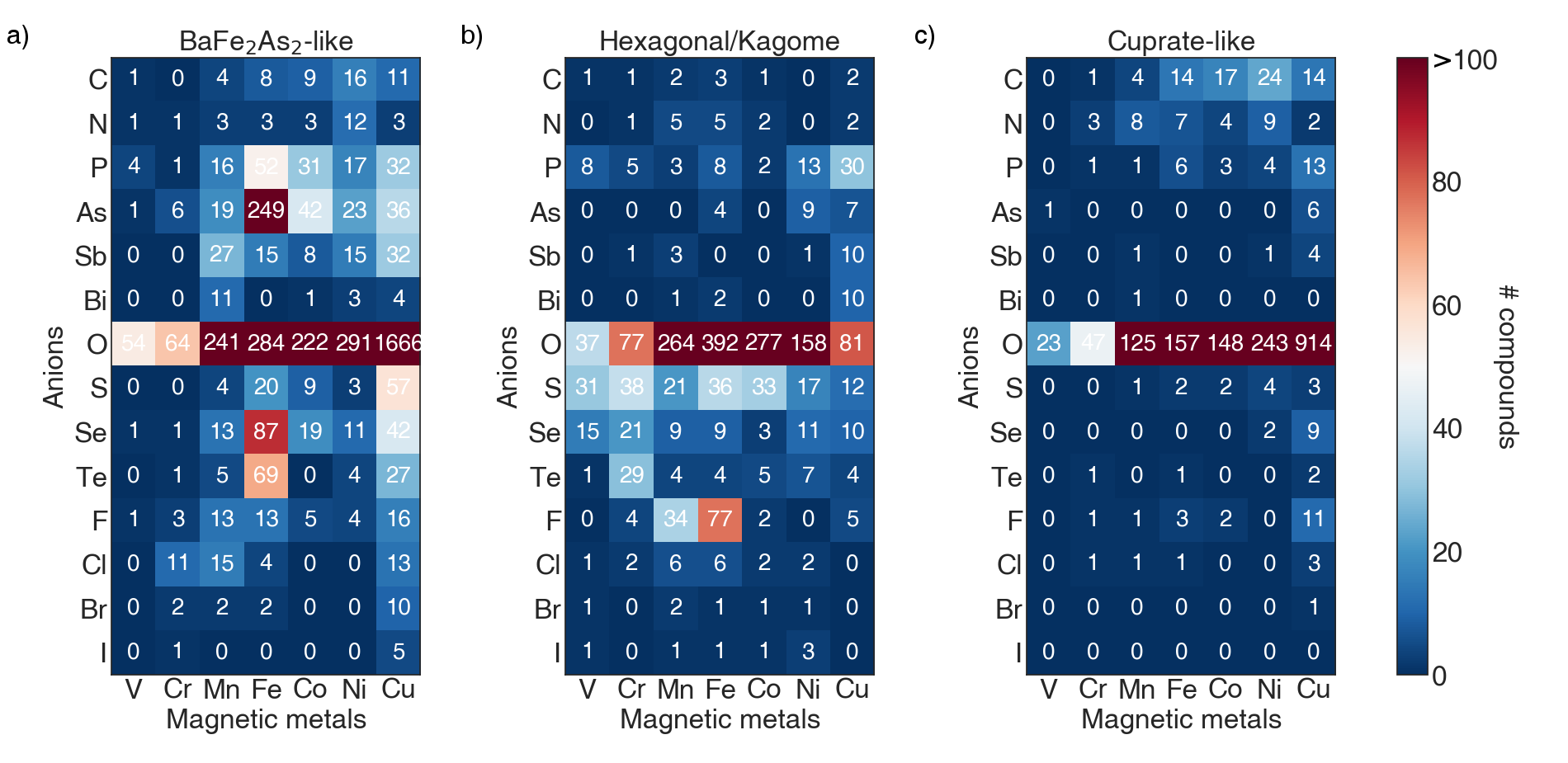} \\
\caption{The distribution of compounds over different anions and magnetic cations is shown for magnetically-2D compounds having a $\Delta_\text{range} \geq 0.8$~\AA\ that are (a) ``BaFe$_2$As$_2$-like'' with planar magnetic cations are arranged in squares (85-95$^\circ$, cation self-coordination 4); (b) frustrated with planar magnetic cations in triangles (55-65$^\circ$, cation self-coordination 4 or 6); and (c) ``cuprate-like'' with planar magnetic cations 4-coordinated in flat sheets by anions.}
\label{fig:cation_anion_distribution}
\end{figure*}

In addition to grouping well-known classes of materials, we wish to uncover compounds that exhibit underrepresented connectivities and structure types. 
For example, it comes as no surprise in the crystal system distribution shown in Figure \ref{fig:system_stype}(a) that tetragonal compounds are often magnetically 2D, but it is perhaps counterintuitive that a few cubic compounds that are actually magnetically-1D, such as $Pm\overline{3}$ YBa$_4$Cu$_3$O$_9$.\cite{abbattista_comprehensive_1989}
The numbers of compounds populating each structure type are shown in Figure \ref{fig:system_stype}(b). The shape of these distributions can be considered a combination of the natural tendency of elements to form similar structures and reporting bias in an experimental database. Most of the compounds lie in a few well-known structure types. The effectiveness of the algorithm lies in recognizing some of the lesser-known structures present in the tail of the distributions.

We can further examine the more ``rare'' compounds by evaluating bonding networks.
Filtering compounds that are magnetically-2D with $\Delta_\text{range} \geq 0.8$~\AA, cation-anion coordination of 4 and cation-cation-cation angle between 85$^\circ$-95$^\circ$ (``BaFe$_2$As$_2$-like'') gives us the distribution of compounds organized by respective anions and magnetic cations shown in Figure \ref{fig:cation_anion_distribution}(a). 
We can clearly see that compounds containing Cu-O, Fe-Se, Fe-Te, and Fe-As are prevalent, as expected.
The ``rare'' entries include Cs(VF$_4$), the only square-lattice vanadium fluoride, which is a low temperature antiferromagnet.\cite{Hidaka1989} 
Given the similarity in cation-anion coordination in Figure \ref{fig:coordination}(a), it becomes clear that an opportunity arises to form V analogs of the more numerous magnetically-square-planar Mn fluorides, for example to produce NaRb$_2$V$_3$F$_{12}$, Sr$_2$VO$_3$F, or LiVF$_4\cdot$H$_2$O (corresponding to Mn-based ICSD codes 83871, 291640, and 417512, respectively).\cite{englich_jahn-teller_1997,su_high-pressure_2016,massa_jahn-teller-ordnung_2007}
Similar logic can be applied to the landscape around Rb$_2$CrCl$_2$I$_2$, which is the only such compound in the Cr-I system.\cite{fyne_powder_1985}

Searching for hexagonal/kagom\'{e} ordering (metal-metal-metal angle 55$^\circ$-65$^\circ$) again shows a high prevalence for oxides in Figure \ref{fig:cation_anion_distribution}(b). Delafossites and LuMnO$_3$ structure types dominate the Cu-O population along with some of the known quantum spin liquids. As mentioned earlier, the Cu-F set contains all 4 of the $A_2$BCu$_3$F$_{12}$ quantum spin liquids alongside BaCu(CO$_3$)F$_2$. 
There are, however, fewer fluorides for Cu than for other cations, and no entries in the set of Cu-containing chlorides, which should be oxidizing enough to form Cu$^{2+}$.
There may then arise opportunities to create new Cu-containing analogs of these halides, for example by substitution into Na$_2$Mn$_3$Cl$_8$.\cite{loon_crystal_1975}


A third exemplary search, for square-planar coordination of the cation to the anion (a truly planar layer) is shown in Figure \ref{fig:cation_anion_distribution}(c).
Again, cuprate oxides are strongly represented.
But more importantly, we quickly gain a picture of when the other anions do form such planar lattices.
For V-containing compounds, only As is known as a suitable anion (tetragonal Zr$_{1.43}$V$_{0.57}$As has La$_2$Sb structure type),\cite{dashjav_crystal_2002} but the existence of Mn-based fluorides, chlorides, and phosphides presents an opportunity to create square lattice materials with a hitherto-unexplored V cation. 
Similar spaces can be investigated by analogy (Co tellurides, chlorides, bromides, etc.).


\subsection{Quantifying intermediate dimensionality}

\begin{figure}
\centering\includegraphics[width=0.85\columnwidth]{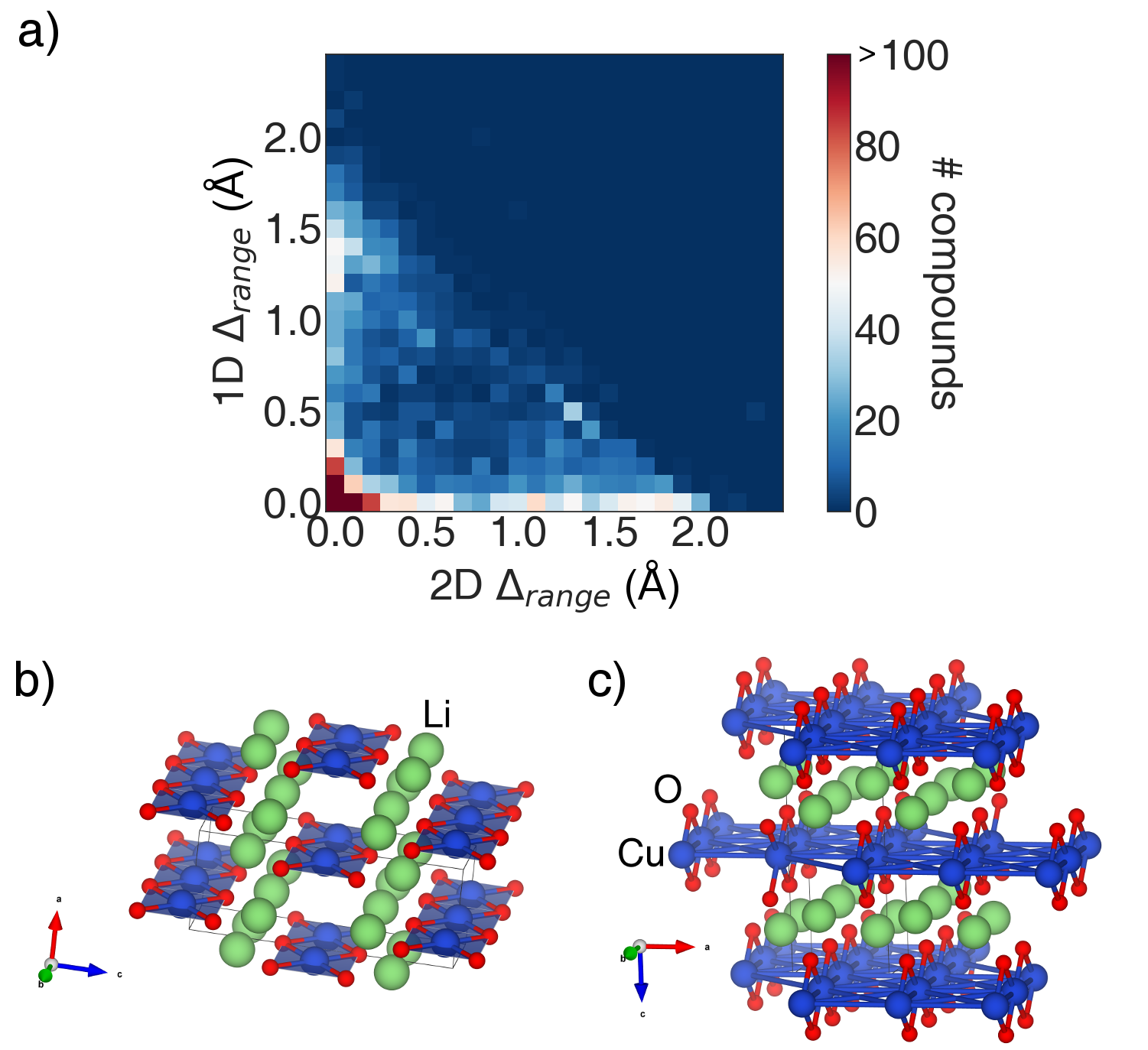} \\
\caption{The quantitative predictor $\Delta_\text{range}$ can give intermediate values in (a), which shows the distribution of compounds having both magnetic 1D and 2D character. Such a compound near the middle of the plot in (a) is Li$_2$CuO$_2$, which can be depicted as (b) magnetically 1D and (c) magnetically 2D.} 
\label{fig:1D_2D_intermediate}
\end{figure}

One key utility of a quantitative dimensionality predictor is its ability to identify compounds that may have complex coupling between spin and lattice degrees of freedom.
Compounds with intermediate $\Delta_\text{range}$ can be examined to identify cases where 1D and 2D behavior compete, which could lead to strong dependence of spin susceptibility to chemical and mechanical pressure. 
The plot shown in Figure \ref{fig:1D_2D_intermediate}(a) represents a histogram of $\Delta_\text{range}$ values of compounds that return nonzero $\Delta_\text{range}$ for both 1D and 2D magnetism. 
Most compounds in this set are clustered around the axes, but many have equally significant 1D and 2D character.
For example, Li$_2$CuO$_2$  
has 1-D and 2-D $\Delta_\text{range}$ of 0.7 and 1.0~\AA, respectively,
and the two corresponding views of the structure are shown in Figure \ref{fig:1D_2D_intermediate}(b,c).
It has a similar structure as CuGeO$_3$, which is a quasi-1D spin Peierls compound with edge-sharing CuO$_4$ square planar units. 
The Cu ions within the 1D chains in this compound are coupled ferromagnetically and the inter-chain interactions are antiferromagnetic.\cite{Sapina1990} 
The N\'{e}el temperature of Li$_2$CuO$_2$ is around 10~K, but inelastic neutron scattering has shown that interchain interactions are strong, and the magnetic ordering is near an instability.\cite{boehm_competing_1998}
Under 5.4~GPa applied pressure, Li$_2$CuO$_2$ transforms into a monoclinic structure which has not yet been characterized magnetically but has shorter interchain distances.\cite{you_high_2009}
A less-explored compound is  Sr$_2$Mn$_2$O$_4$Se, which
has 1D and 2D $\Delta_\text{range}$ of 0.8 and 1.1~\AA, respectively. 
This oxyselenide compound 
contains octahedral Mn$^{3+}$ ions which are arranged in corrugated 1D chains.
Sr$_2$Mn$_2$O$_4$Se orders antiferromagnetically at 160 K, followed by a second antiferromagnetic ordering transition at 126 K.\cite{Free2012} 
The interchain ordering in Sr$_2$Mn$_2$O$_4$Se determines the relatively high 160~K N\'{e}el temperature, which can be contrasted to more strongly 1D (but not corrugated) Mn-containing oxides, such as the linear compounds K$_5$Mn$_3$O$_6$ or Cs$_4$Mn$_3$O$_6$, which have very strong intrachain interactions, evidenced by their non-Curie-Weiss susceptibility, but no long-range ordering down to liquid helium temperatures.\cite{pfeiffer_chain_2010,nuss_k5mn3o6_2015}
Other compounds in our dataset fall into this intermediate regime and have yet to be explored in detail.

\section{Conclusions} 
In the course of developing a quantitative geometric predictor for magnetic dimensionality $\Delta_\text{range}$, the evaluated bonding networks provide a robust method for filtering known compounds and identifying uncharted coordinations.
This method groups compounds with similar motifs and coordination, even when such associations span many space groups or crystal systems. 
Compounds with intermediate $\Delta_\text{range}$ comprise an interesting set of materials that may host complex spin-lattice coupling.
The dataset, search algorithm code, and a searchable frontend are freely available.

\section{Acknowledgments}

We thank J. N. Eckstein for helpful discussions. This work was supported by the Center for Emergent Superconductivity, an Energy Frontier Research Center funded by the United States Department of Energy, Office of Basic Energy Sciences.

\bibliography{magnetic2d}

\end{document}